# Tunable Multi-Peak Perfect Absorbers Based on Borophene for High-Performance Near-Infrared Refractive Index Sensing

Ruda Jian[1,#], Shiwen Wu[1,#], Bo Zhao[2], Guoping Xiong[1,*]

[1]Department of Mechanical Engineering, The University of Texas at Dallas, Richardson, TX 75080, United States.

[2]Department of Mechanical Engineering, University of Houston, Houston, TX 77004, United States.

# These authors contribute equally to this work.

*Corresponding author. Email: Guoping.Xiong@UTDallas.edu

**Abstract**

Borophene has recently attracted significant attention as an emerging two-dimensional monoelemental material for refractive index sensing because of its ultra-high surface-to-volume ratio and outstanding surface sensitivity. However, current research mainly focuses on designing borophene-based sensors with single-peak absorption, which lacks reliability compared to the performance of multi-peak sensors. In this paper, high-performance borophene refractive index sensors with multiple nearly perfect absorption peaks are proposed. The geometric parameters of the metadevices are optimized using finite-difference time-domain (FDTD) simulations to obtain three strong absorption peaks in the near-infrared (near-IR) regime with intensities of 99.72%, 99.18%, and 98.13%. Further calculations show that the sensitivities of the three strong peaks reach 339.1 nm shift per refractive index units (nm/RIU), 410.1 nm/RIU, and 738.1 nm/RIU, and their corresponding figure of merits (FOMs) reach up to 15.41/RIU, 14.65/RIU and 10.85/RIU, respectively, exhibiting great potential for near-IR refractive index sensing. Moreover, the three absorption peaks can be easily tuned by adjusting the carrier density of borophene, which can be realized by applying different external electric bias voltages. These results can provide theoretical guidance for the development of multi-peak near-IR dynamic integrated photonic devices.

**Keywords**  Borophene · Multi-peak · Perfect absorption ·  Refractive index sensor

**Introduction**

As a new member of the two-dimensional (2D) monoelemental material (Xene) family, borophene has recently attracted tremendous attention because of its high thermal conductivity, excellent mechanical and electronic transport properties, and anisotropic optical behavior [1–4]. Similar to other 2D materials [5–8], borophene exhibits great potential for refractive index sensing because of its ultra-high surface-to-volume ratio and outstanding surface sensitivity [7–9]. In particular, borophene possesses a high carrier density (~$10^{19}$ m$^{-2}$) [13,14], several orders of magnitude higher than those of other 2D materials such as graphene and black phosphorus (~$10^{17}$ m$^{-2}$) [12,13], enabling borophene-based plasmonic devices to work in the visible and near-infrared (near-IR) regimes. In contrast to visible or near-IR sensors based on noble metals [17,18], the working wavelength of borophene devices can be easily tuned by applying external electric bias voltages to control its carrier density [19,20], making borophene-based sensors potentially suitable to operate in a broad wavelength range. Meanwhile, borophene devices possess unique advantages of 2D resonators such as high mode volume and strong light-confinement in the atomic scale [21], resulting in high detection sensitivity.

Although many refractive index sensors based on 2D materials have been widely reported, most of them are based on a single-peak absorption mechanism [9,22,23]. Compared to refractive index sensors relying only on single absorption peak, sensors with multiple absorption peaks can be used to monitor multiple modes of interest, and these multiple absorption peaks can generate more sensing information, thus reducing the error and improving the reliability of data [12,24,25]. For instance, refractive index sensors based on patterned graphene or gold layers exhibit three absorption peaks in the mid-IR regime [11,26] and can provide reliable sensing data associated with multiple vibrational signatures of targeted molecules. To date, borophene refractive index sensors with multiple absorption peaks have not been reported.

In this letter, we propose a design of multi-peak borophene-based metadevices for efficient refractive index sensing applications. Our design can achieve three nearly perfect absorption peaks in the near-IR regime, resulting in excellent sensing performance that is comparable to or even better than that of the single-peak refractive index sensors in existing work. Moreover, these absorption peaks can be easily tuned by changing the carrier density of borophene, making the proposed borophene sensors highly promising for operations within a broad wavelength range.

**Structure and Computational Method**

The schematic diagrams of the borophene devices are shown in Figs. 1a-b. Fig. 1a displays the global view of a unit cell in a typical metadevice, in which a borophene grating layer lies on a dielectric layer with a refractive index of 1.4, and a gold film is employed as a reflective layer placed at the bottom of the absorber. Here, the thickness of borophene $d_B$ = 0.3 nm is adopted to simulate the borophene monolayer [9,27]. The thicknesses of the dielectric layer $d_1$ and gold film $d_2$ are set to be 150 nm and 100 nm, respectively. As shown in the top view of a unit cell in the metadevice (Fig. 1b), the patterned borophene layer comprises of two symmetrical nano-ribbons and a nano-rectangle. $p$ represents a single period of the unit cell. $l$ and $w$ represent the length and width of the nano-ribbons, respectively. $g$ and $h$ represent the $x$ span and $y$ span of the nano-rectangle, respectively. *s and m* represent the distance between the nano-ribbon and the nano-rectangle along $y$ direction and the distance between the nano-rectangle to the edge along the $x$ direction, respectively. $j$ and $k$ represent the distance between the nano-ribbons and the edge along the $x$ direction and $y$ direction, respectively. The distances (denoted as $s$) between the nano-rectangle and both nano-ribbons are kept the same.

The interactions between incident light and borophene metadevices are simulated by commercial finite-difference time-domain (FDTD) software (Lumerical 2021 R2.2) with a three-dimensional model constructed in the FDTD solver. Periodic boundary conditions are employed in both $x$ and $y$ directions to mimic the periodic structure of the borophene pattern, and perfect matching layer (PML) boundary condition is employed in the $z$ direction. Two power monitors are placed 2 μm above and under the borophene metadevice to collect the reflectivity ($R$) and transmissivity ($T$), respectively. The absorptivity ($A$) spectrum can thus be calculated from the corresponding reflectivity and transmissivity spectra as $A = 1 - R - T$. Note that $T$ is zero since the structure is opaque (Supplementary Fig. S1). In all simulations, mesh sizes of 1 nm in the $x$ and $y$ directions, and 0.15 nm in the $z$ direction are used. Optical parameters of gold are taken from Ref. [28]. A Drude model is employed to calculate the anisotropic optical conductivity of borophene [29]:

$$\sigma_{jj}(\omega) = \frac{iD_j}{\pi(\omega + i\tau^{-1})}, D_j = \frac{\pi e^2 n_s}{m_j} \quad (1)$$

where $j$ indicates the direction of optical axes of borophene crystal and can be chosen as $x$ or $y$. $\omega$ is the angular frequency of light, and $\tau$ is the electron relaxation time and set as 65 fs. $D_j$ represents the Drude weight, where $e$ is the electron charge, and $n_s$ is the carrier density and is initially set as $4.3 \times 10^{19}$ m$^{-2}$. $m_j$ is the effective electron mass in two crystal axis directions. Here, $m_x = 1.4\ m_0$ and $m_y = 5.2\ m_0$, where $m_0$ represents the rest mass of electrons. The effective permittivity of the borophene monolayer can thus be derived from the surface conductivity along each direction as [29]:

$$\varepsilon_{jj} = \varepsilon_r + \frac{i\sigma_{jj}}{\varepsilon_0 \omega d_B} \quad (2)$$

where $\varepsilon_r = 11$ is the relative permittivity of borophene [29], $\varepsilon_0$ is the permittivity of free space, and $d_B$ is the thickness of borophene. Similar to other 2D materials, the z-component permittivity of borophene is set as $\varepsilon_{zz} = \varepsilon_r$ [19,30].

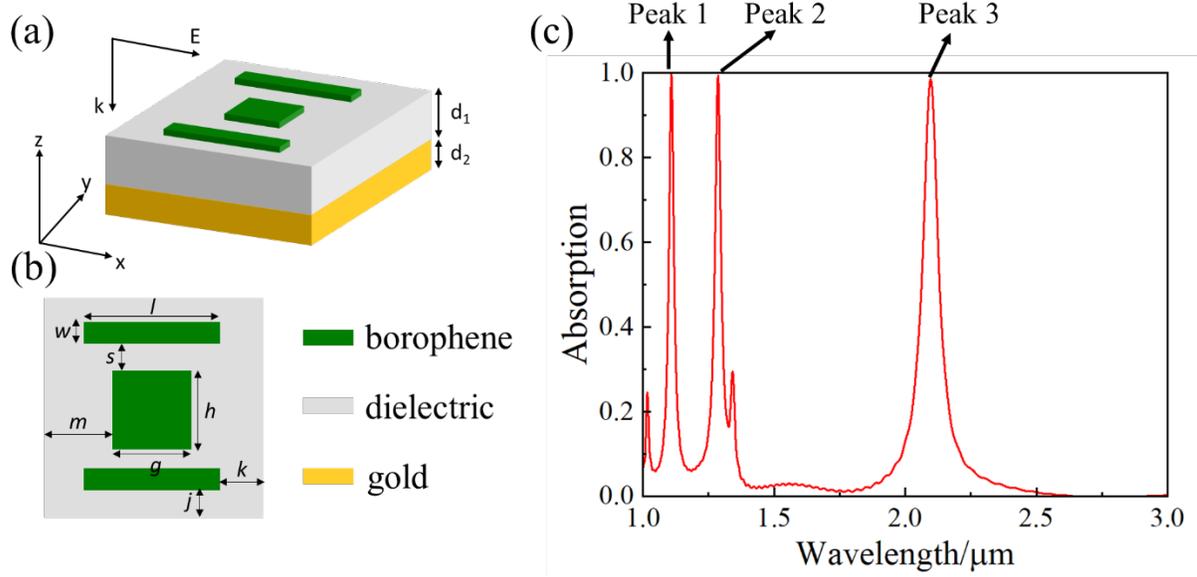

Fig. 1. (a) Global view and (b) top view of a unit cell in a typical borophene metadevice. (c) Absorption spectrum of a borophene metadevice under a normal incidence of light, when the geometric parameters of the patterned borophene layer are set as $l$ = 55 nm, $w$ = 10 nm, g = 30 nm, $h$ = 30 nm, and s = 20 nm. The thicknesses of borophene, dielectric layer, and gold substrate are 0.3 nm, 150 nm and 100 nm, respectively.

**Results and Discussion**

Absorption peaks with high intensities are preferred to achieve excellent sensing performance for refractive index sensors [31]. The absorption spectra of the borophene metadevices under a normal incidence of light within the wavelength range of 1 to 3 µm are calculated. As shown in Fig. 1c, three nearly perfect absorption peaks located at $\lambda_1$ = 1.108 µm, $\lambda_2$ = 1.286 µm, and $\lambda_3$ = 2.098 µm, corresponding to absorptivities of 99.72%, 99.18%, and 98.13%, respectively, are obtained after optimizations of the geometric parameters ($l$ = 55 nm, $w$ = 10 nm, $g$ = 30 nm, $h$ =

30 nm, and $s$ = 20 nm) of a typical borophene metadevice. Details of structural optimizations are shown in Supplementary Fig. S2.

To reveal the origin of the multi-peak absorption of the borophene metadevice, we further calculate the electric field distributions within the patterned borophene layer at the wavelengths of the absorption peaks (Fig. 2). At $\lambda_1$ = 1.108 μm, electric field enhancement $|E/E_0|$ primarily occurs at the edges of the nano-rectangle along the $y$ axis (Fig. 2a). The $z$-component of the electric field ($E_z$) distribution shown in Fig. 2b indicates that strong $E_z$ can be observed at the edges of the nano-rectangle along the $y$ axis, along with weaker $E_z$ distributed at the corners of the nano-rectangle. Apparent positive and negative dipoles are located at those positions (Supplementary Fig. S3), which can be attributed to the localized surface plasmon resonance effect [19,32] of the borophene nano-rectangle. Meanwhile, strong electric field localization can be observed in the R$_1$ region from the $x$-component of the electric field ($E_x$) distribution at $\lambda_1$ (Fig. 2c), indicating that the middle part of the borophene nano-rectangles provides principal channels for the energy transportation at $\lambda_1$ [19,33]. At $\lambda_2$ = 1.286 μm, electric field enhancement is mainly located at the four corners of the borophene nano-rectangle (Fig. 2d), which may arise from the corner effect due to the sharpness in shape [19]. The $E_z$ distribution in Fig. 2e and Supplementary Fig. S3 shows that dipoles can also be observed at the four corners at $\lambda_2$, with inverted '±' signs compared to the ones at the four corners at $\lambda_1$. Moreover, as shown in the $E_x$ distribution in Fig. 2f, strong electric field localization in the R$_2$ region illustrates that the edges of the borophene nano-rectangle along x axis are the primary channels for the energy transportation at $\lambda_2$. The results indicate that interactions between the borophene nano-rectangle and light lead to the formation of two strong absorption peaks at $\lambda_1$ and $\lambda_2$. While at $\lambda_3$ = 2.098 μm, the electric field enhancement is mainly distributed along the edges of the borophene nano-ribbons (Fig. 2g). Detailed $E_z$ distribution (Fig. 2h and Supplementary Fig.

S3) shows that positive and negative dipoles are observed at the edges of the nano-ribbons, and the *x*-component electric field is primarily located at the edges of the nano-ribbons (Fig. 2i), illustrating that plasmon hybridization between borophene nano-ribbons in adjacent unit cells along the *x* axis may have led to the formation of the absorption peak at $\lambda_3$ [32,34].

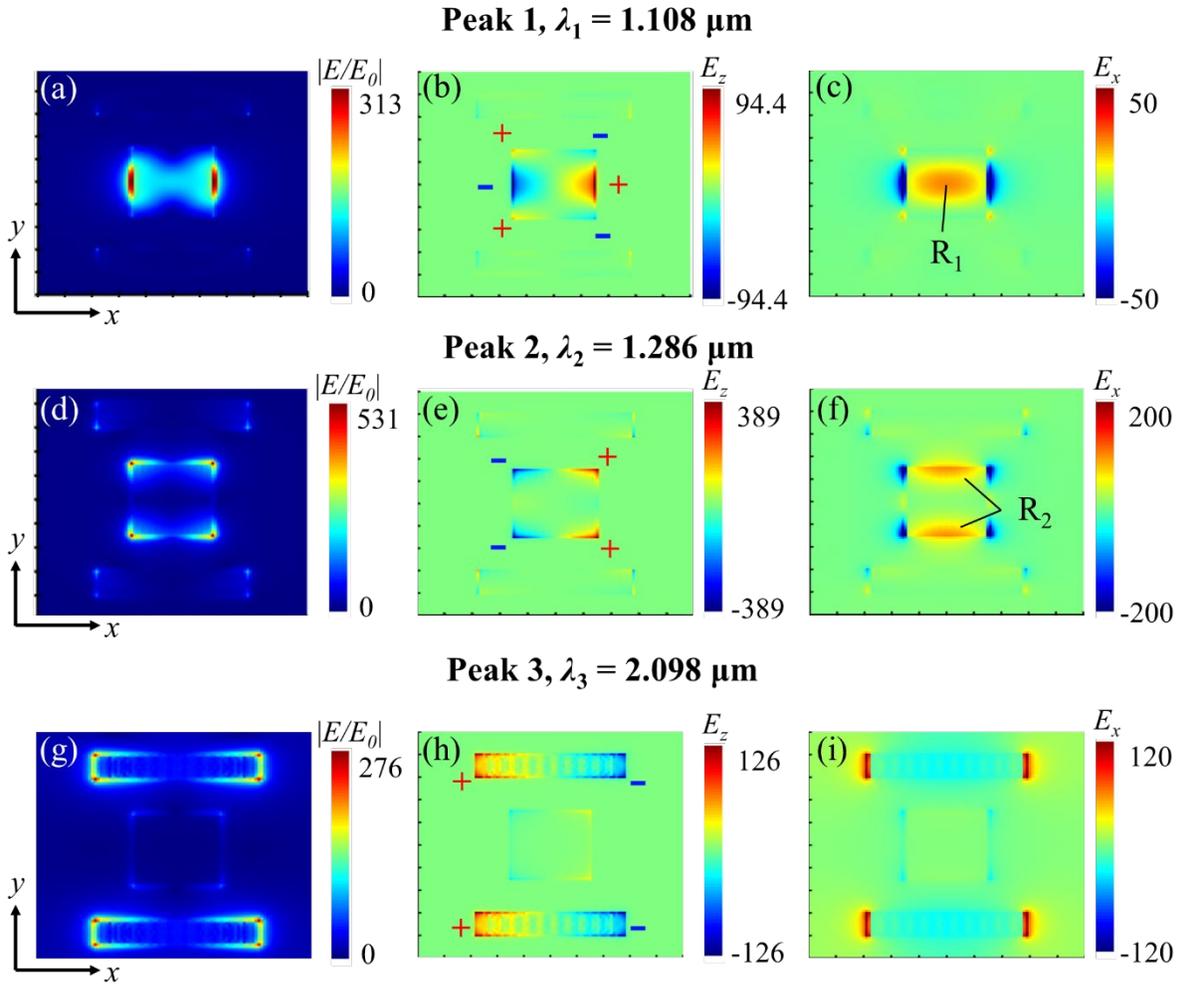

Fig. 2. Distributions of (a) the electric field enhancement $|E/E_0|$, (b) the *z*-component electric field $E_z$, and (c) the *x*-component electric field $E_x$ within the patterned borophene layer in the *x-y* plane ($z$ = 350 nm) at the absorption peak $\lambda_1$ = 1.108 μm. Distributions of (d) the electric field enhancement $|E/E_0|$, (e) the *z*-component electric field $E_z$, and (f) the *x*-component electric field $E_x$ within the patterned borophene layer in the *x-y* plane ($z$ = 350 nm) at the absorption peak $\lambda_2$ = 1.286 μm. Distributions of (g) the electric field enhancement $|E/E_0|$, (h) the *z*-component electric field $E_z$,

and (i) the x-component electric field $E_x$ within the patterned borophene layer in the x-y plane ($z$ = 350 nm) at the absorption peak $\lambda_3$ = 2.098 μm.

Next, we demonstrate the outstanding refractive index sensing performance of the multi-peak borophene metadevices. Fig. 3a shows the absorption spectra of an optimized borophene metadevice ($l$ = 55 nm, $w$ = 10 nm, $g$ = 30 nm, $h$ = 30 nm, and $s$ = 20 nm) under different environmental refractive indices ($n_{env}$). As $n_{env}$ increases, all of the three resonant peaks shift to longer wavelengths. The sensitivity (S) of the refractive index sensor can thus be calculated by $S = \Delta\lambda/\Delta n_{env}$, where $\Delta\lambda$ represents the wavelength shift of the resonant peak, and $\Delta n_{env}$ represents the change in the refractive index of the environment [9]. Figure of merit (FOM), indicating the optical resolution of the refractive index sensor [35], is calculated by FOM = S/FWHM, where FWHM represents the full width at half maximum of a resonance peak. Fig. 3b shows the resonant wavelengths of the peaks in the absorption spectra of the metadevice under different environmental refractive indices. The sensitivities at the three resonant wavelength peaks are calculated to be 339.1, 410.1, and 738.1 nm shift per refractive index units (nm/RIU), and the corresponding FOMs are calculated to be 15.41/RIU, 14.65/RIU and 10.85/RIU, respectively. Moreover, the normalized sensitivity $S'$ is calculated to compare the performance of various sensors by: $S' = S/\lambda_{resonant}$, where $\lambda_{resonant}$ represents the resonant wavelength [9,31]. The calculated normalized sensitivities of the absorption peak1, peak 2 and peak 3 are 0.31/RIU, 0.32/RIU and 0.35/RIU, respectively, indicating the excellent sensing performance of the proposed multi-peak borophene refractive index sensors, comparable to or even higher than that of single-peak sensors reported in prior work (Supplementary Table S1).

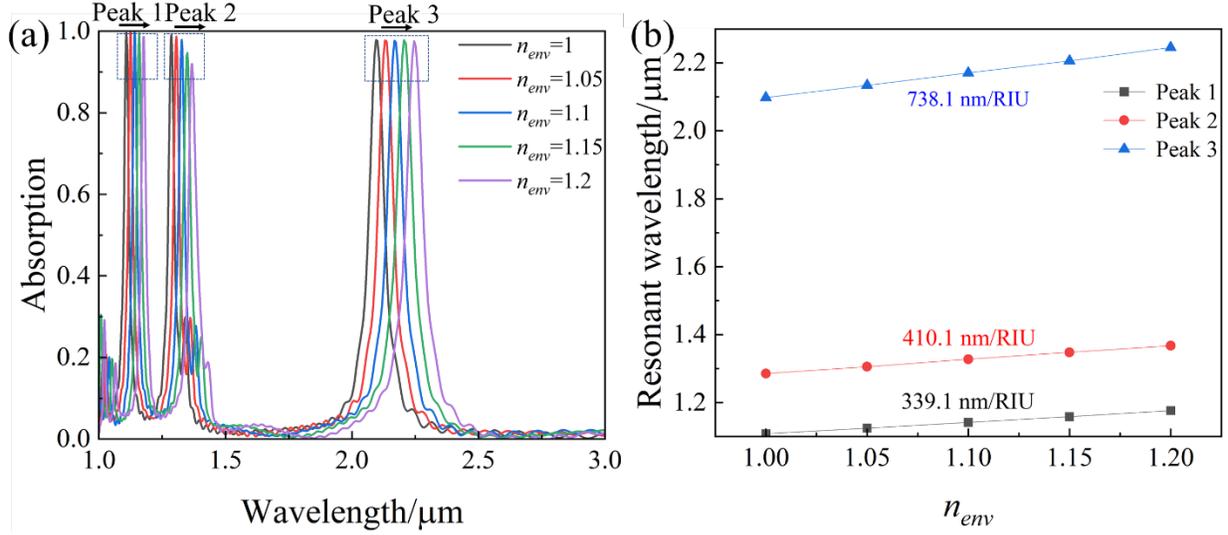

Fig. 3. (a) Absorption spectra of the borophene metadevice as the environmental refractive index $n_{env}$ changes from 1 to 1.2. (b) Resonant wavelengths of the borophene metadevice under different $n_{env}$.

Compared to noble metals, the optical properties of borophene can be easily tuned by applying different external electric bias voltages to adjust the carrier density [14,36]. Fig. 4a shows the absorption spectra of the third absorption peak of the borophene sensors with different carrier densities ($n_s$) ranging from $3.3\times10^{19}$ m$^{-2}$ to $5.3\times10^{19}$ m$^{-2}$, while those of the first and the second peaks are shown in the Supplementary Fig. S4. When $n_s$ increases, all the three absorption peaks are blue-shifted, and the *FWHM* of each peak narrows gradually. We further reveal the effect of $n_s$ on the sensing performance of each absorption peak. As shown in Figs. 4b-d, the resonant wavelengths of the three absorption peaks versus $n_{env}$ for the sensors with different $n_s$ are plotted. When $n_{env}$ increases, all the resonant peaks shift to longer wavelengths. Moreover, the sensitivity calculated at each resonant peak keeps increasing as $n_s$ decreases, indicating the highly tunable properties of the borophene sensors. For instance, the sensitivity of the borophene metadevice reaches up to 856.0 nm/RIU when the carrier density of borophene is adjusted to $3.3\times10^{19}$ m$^{-2}$ (Fig. 4d).

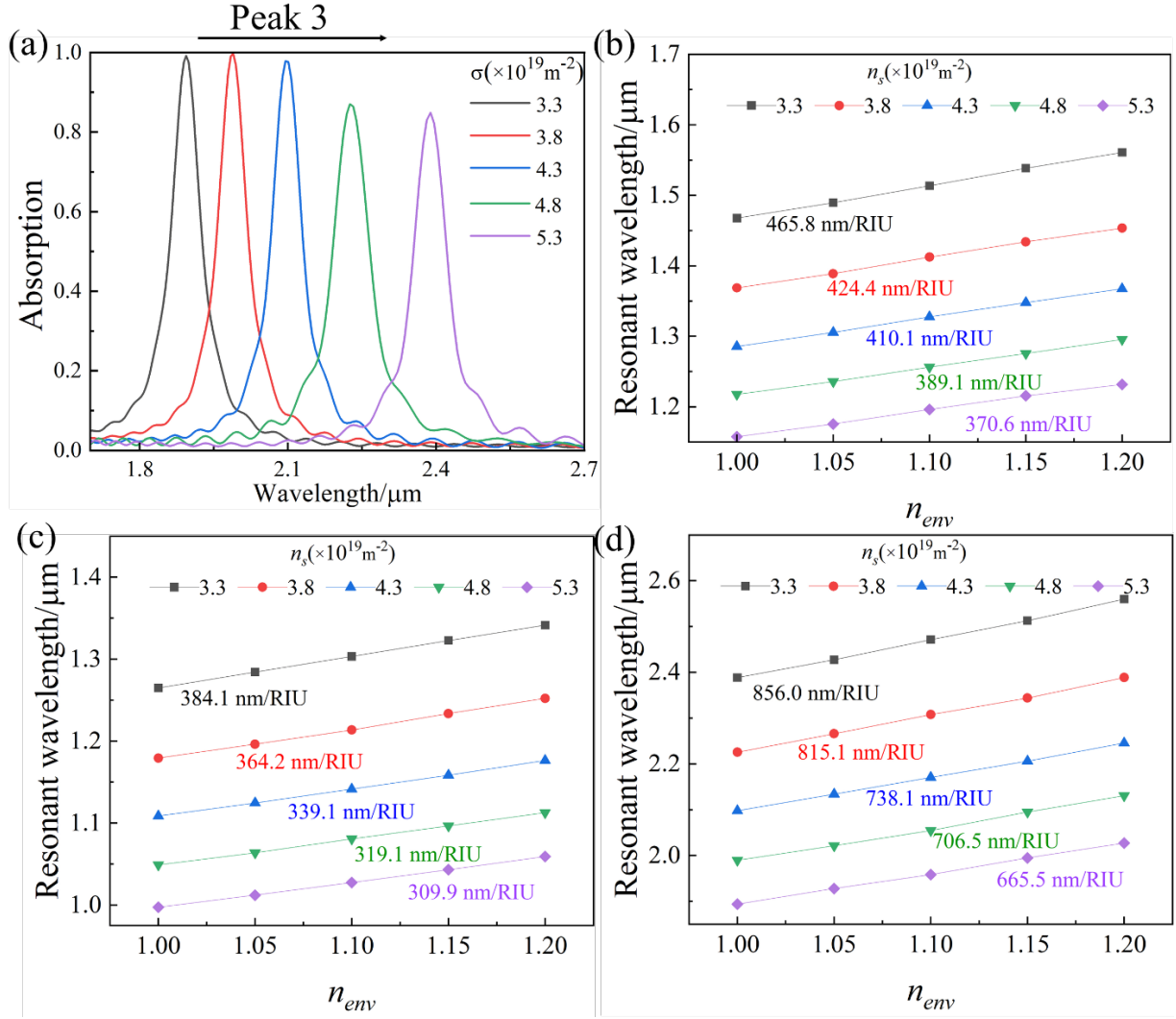

Fig. 4. (a) Absorption spectra of the third absorption peak of the borophene metadevice with different carrier densities ($n_s$) of borophene. (b) The resonant wavelengths of absorption peak 1 as functions of $n_{env}$ under different $n_s$. (c) The resonant wavelengths of absorption peak 2 as functions of $n_{env}$ under different $n_s$. (d) The resonant wavelengths of absorption peak 3 as functions of $n_{env}$ under different $n_s$.

## Conclusion

In summary, we have designed efficient borophene-based refractive index sensors with multiple nearly perfect absorption peaks in the near-IR regime. After optimizing the geometry parameters of the borophene metadevices, three strong absorption peaks with intensities of 99.72%, 99.18%, and 98.13% are achieved. Detailed analyses show that the sensitivities calculated at the peaks reach 339.1 nm/RIU, 410.1 nm/RIU, and 738.1 nm/RIU, and corresponding FOM reaches 15.41/RIU, 14.65/RIU and 10.85/RIU, respectively, exhibiting excellent sensing performance that is comparable to or even higher than that of single-peak refractive index sensors in prior work. In addition, we show that the working wavelength and sensing performance of the borophene sensors can be easily tuned by adjusting the carrier density of borophene. This study demonstrates the potential applications of borophene in the next-generation tunable sensing systems operating in the near-IR regime and provides useful insights into the design of multi-peak borophene-based optoelectronic and photonic nanodevices.

**Funding information**

The University of Texas at Dallas startup fund; National Science Foundation (Grant No. CBET-1937949).

**Disclosures**

The authors declare no conflicts of interest.

**Data availability**

The datasets analyzed during the current study are available from the corresponding author on reasonable request.

**Supplemental document**

See Supplementary materials for supporting content.